# The Low Energy Beam Transport System of the New GSI High Current Injector

L. Dahl, P. Spädtke, GSI Darmstadt, Germany

*Abstract*

The UNILAC was improved for high current performance by replacing the Widerøe prestripper accelerator by an RFQ and an IH-type DTL. In addition, one of two ion source terminals was equipped with high current sources of MUCIS- and MEVVA-type. Therefore, a redesign of the LEBT from the ion source to the RFQ entrance was necessary. The new LEBT was installed at the beginning of 1999. The commissioning was carried mainly with a high intense argon beam. The achieved 17 emA $Ar^+$ are 6 emA above the RFQ design intensity. The expected performance of the new LEBT was achieved: full transmission for high current beams, preservation of brilliance along the beam line, isotope separation for all elements, transverse phase space matching to the RFQ linac, and macropulse shaping. In particular, the high degree of space charge compensation was confirmed.

## 1 INTRODUCTION

To fill up the heavy ion synchrotron SIS to its space charge limits for all ion species, prestripper intensities of $I = 0.25 \cdot A/q$ (emA) for mass over charge ratios up to 65 are neccessary. For this goal the Widerøe accelerator was substituted by a high current IH-RFQ and two IH-type cavities working at 36 MHz [1]. To compensate the calculated transmission loss in the RFQ the beam current in the LEBT was even required 10 % higher, in case of the design ion $U^{4+}$ the beam current to be transported amounts to 17 emA.

Performance goals for the new LEBT were loss free beam transport from the new source types MUCIS and MEVVA to the RFQ, avoiding of emittance growth, mass resolution $m/\Delta m = 220$ of the magnet spectrometer, macropulse shaping with rise and fall times $\Delta t \leq 0.5$ µs and exact transverse beam emittance matching to the RFQ acceptance. Furthermore, the LEBT from the switching magnet downstream was required to be 50 Hz pulseable for two ion species operation of the high current injector.

## 2 DESIGN OF THE HIGH CURRENT LEBT

Because the high current sources were installed years before, many high intense beam investigations were carried out at the former beam transport system to get a basis for the ion optical design of the new LEBT [2]. It turned out that there is no evident emittance growth and the beam behavior agrees very well with zero-current simulations. All measurements indicated a space charge compensation by electrons out of the residual gas of at least 95 %. Therefore, the existing LEBT down to the switching magnet was evaluated as basically useful for the transport of high intense beams. It was improved by the introduction of steering magnets near the ion source and an extension of a magnetic quadrupole doublet to a triplet for a flattened dispersion trajectory. Due to the longer new accelerator tanks, the downstream beam line was significantly shortened. It was redesigned also for better dispersion quality, and to perform the transverse beam matching to the narrow RFQ entrance (Fig.1).

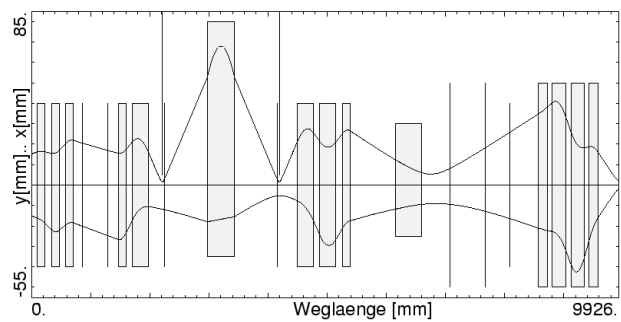

Fig. 2 Transverse envelope of a zero current beam

Fig. 2 shows the transverse beam envelope calculated by the ellipse transformation code MIRKO based on emittance measurements directly behind the acceleration column of the ion source terminal. The 77.5° magnet spectrometer provides a mass resolution of $m/\Delta m = 220$. This allows the isotope separation for all elements and reduces remaining space charge forces in the following system. This LEBT is achromatic only in the way that particles of different energies coincide in a focus at the RFQ entrance but with different angles. Investigations carried out with the multi-particle code PARMT, proved a possible momentum spread of only $\pm 5 \cdot 10^{-4}$. This value was considered for the new high voltage power supply for preacceleration to an energy of 2.2 keV/u. Beam transport is limited to a space charge analogous to an effective beam current of 0.5 emA only. Within this limit no beam losses occur and the emittance growth stays within the RFQ acceptance.

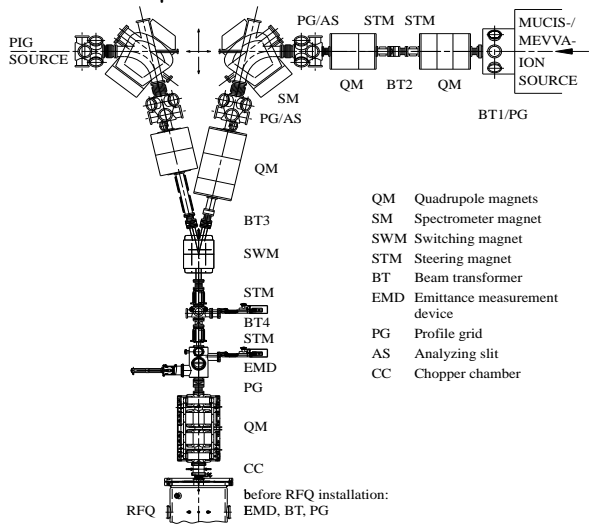

Fig. 1 Mechanical layout of the new LEBT

The high current sources deliver beam pulses of 1-3 ms with a repetition rate of ≤ 17 Hz. The PIG-sources serve the UNILAC experimental area with a second beam of up to 6 ms length and a repetition rate of ≤ 50 Hz. Therefore, time sharing operation with laminated magnets is chosen for the common beam line.

The double waist at the RFQ entrance is performed by four new magnetic quadrupoles with apertures of 100 mm in diameter. Due to the convergence angle of 60 mrad only 240 mm remained for the introduction of an electrostatic transverse chopper for pulse shaping with a conical copper beam dump mounted within a vacuum bellow between the quadrupole magnets and the RFQ entrance. This position was chosen to keep the beam neutralized as close as possible to the RFQ tank.

Fig. 3 presents a schematic view of the chopper area including the deflected beam and equipotential planes of negative voltage calculated by KOBRA3 [3]. The chopper voltage of ± 15 kV rises and drops within $\Delta t \leq 0.5\ \mu s$. A simultaneously pulsed ring electrode (-2 kV) in front of the chopper chamber serves as repeller for the upstream space charge compensation electrons. The potential barrier of a smaller ring electrode with constant voltage (-1 kV) at the end of the beam dump keeps the secondary electrons in the chopper chamber and therefore prevents decompensation of the beam and simultaneously sparking at the RFQ rods. As a further property of the chopper the secondary electrons generated by the primary beam in the beam dump are repelled back.

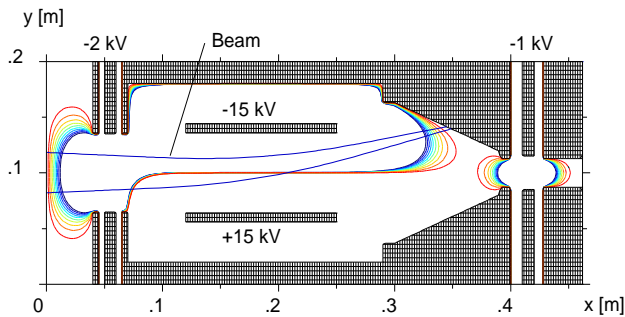

Fig. 3 Equipotential planes with voltages $-10\ V \geq U \geq -100\ V$ and deflected beam in the electrostatic chopper

## 3 COMMISSIONING MEASUREMENTS

For commissioning a beam diagnostic test setup [4] was mounted in place of the RFQ tank, including an emittance measurement device exactly positioned at the RFQ entrance focus, two beam transformers, a profile grid, and a residual gas profile monitor.

First measurements in April 1999 were carried out with an $Ar^+$ beam from the MUCIS source. The analyzed beam current of up to 17 emA was transported without losses to the end of the LEBT. Fig. 4 represents the beam transformer signals. The upper signals are recorded directly behind the ion source (BT1/2). The lower signals show the analyzed beam (BT3/4). The difference in current of around 10 mA results from higher charge states extracted from the ion source. A part of these beam components and eventually also a part of the selected $Ar^+$ beam is lost on the beam tube and causes additional secondary electrons in front of the inflector magnet. For this reason, here the compensation build up time is shorter. To shorten the build up time behind the isotope selection too, a local increase of the vacuum pressure to $1 \cdot 10^{-4}\ Pa$ at the inflector magnet is neccessary.

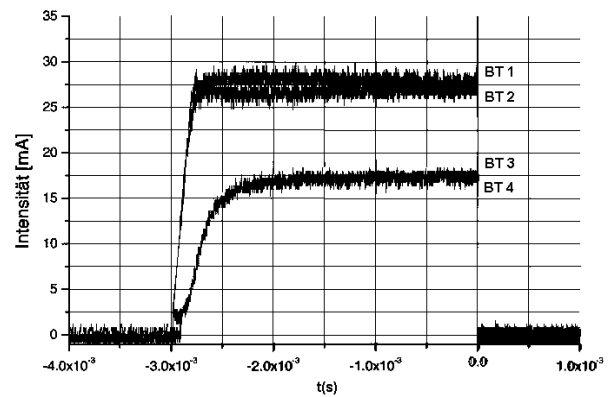

Fig. 4 Beam transformer signals along the beam line.

After further developments the GSI high current sources provide in the meantime raised ion intensities [5] but obviously with changed emittance parameters behind the acceleration gap. These will be measured in January 2001 and consequently the first section of the LEBT might be redesigned for an improved beam transport to the magnet spectrometer.

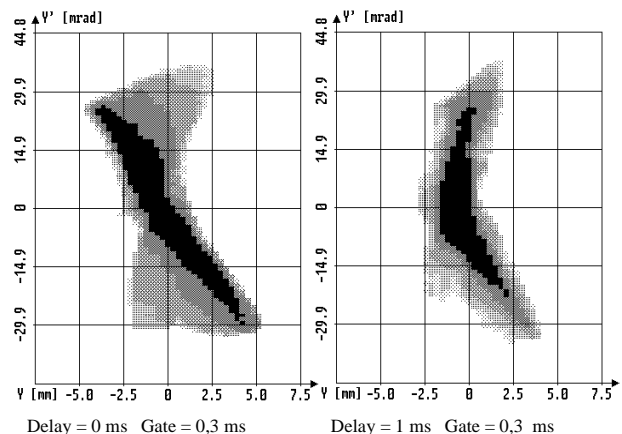

Fig. 5 Different time windows during the macropulse

The transverse emittance of the investigated beam was detected at the end of the LEBT. Fig. 5 shows the effect of defocusing space charge forces during the neutralization build up time. This does not affect the beam of 100 μs pulse length which will be cut out by the chopper in the middle of the flat top for further acceleration. The emittance area depends on the beam current. For $Ar^+$, 7 emA beam current, $\varepsilon_{x,y} = 90\ \pi$ mm mrad was measured, for an intensity of

17 emA, $\varepsilon_{x,y}$ increased up to 130 $\pi$ mm mrad, both values covering 90 % of the total beam. Thus, the transverse beam emittance fits into the RFQ design acceptance of $\alpha_{x,y}$ = 138 $\pi$ mm mrad even for the highest intensities.

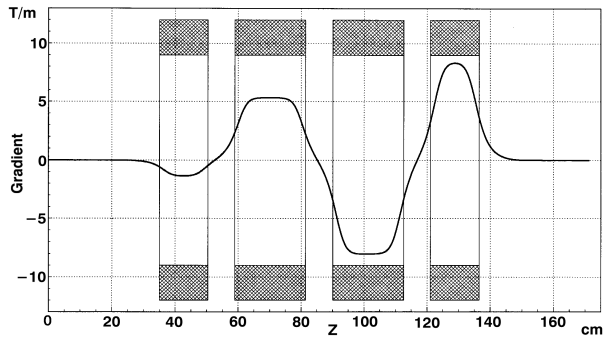

Fig. 6 Measured quadrupole quartet field distribution

The transverse emittance matching to the RFQ acceptance was refined with the multi-particle code DYNAMION [6]. With this code the six dimensional particle motion in the measured magnetic field gradient distribution (see Fig. 6) is calculated by time integration of the equations of motion and therefore the results are most accurate. Related to the first order hard-edged model, the magnet settings turned out to be increased by 6 % for an exact emittance matching to the RFQ. This result was confirmed by measurements.

There are several arguments supporting the assumption of a rather complete space charge compensation in the flat top region of the macropulse. The adjusted quadrupole settings were very close to the zero-current envelope calculation. E.g., an effective space charge corresponding to 1 emA decompensated beam current would cause an increase of the field gradients of 25 % in average. The transverse emittance areas measured behind the switching magnet and at the end of the LEBT were similar. A local increase of the vacuum pressure shortened the pulse rise time but did not influence the flat top currents observed by the beam transformer signals. A hint of a slight non-saturation of the beam with secondary electrons was observed at the sharply focused beam behind the magnet spectrometer. A position of the analyzing slits very close to the beam increased the beam intensity downstream. To preserve the beam current here, the quadrupole doublet gradients in front of had to be raised by less than 3 %.

The experimentally determined dispersion of the spectrometer of 11.46 mm/% verified the theoretical value of D = 12.00 mm/%, sufficient for the separation of Pb isotopes.

After installation and running-in of the RFQ accelerator in May 1999 also capacitive phase probe measurements behind the RFQ tank were available to observe single bunches and the macropulse structure. Fig. 7 shows the macropulse rise time to full intensity of $\Delta t \leq 0.5$ $\mu$s for an $Ar^+$ beam of 9 emA.

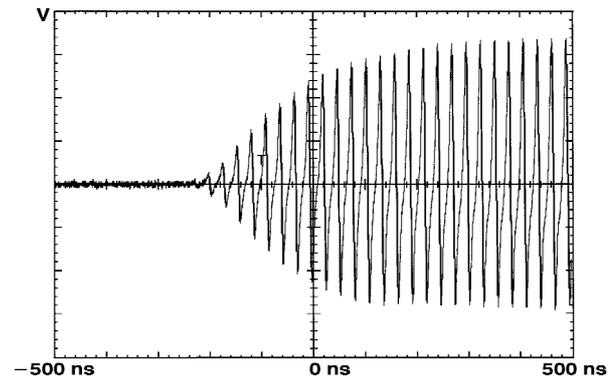

Fig. 7 Beam macropulse rise behind the RFQ

Within this time rf phase and amplitude have to be controlled to stable values. The respective amount of beam energy could cause damages if it is not limited by a short rise time.

Similar investigations were carried out with $^{58}Ni^+$ (4 emA) and $^{238}U^{4+}$ (8 emA) from the MEVVA source. Although the pulses from this source are very noisy and therefore fluctuations of the space charge compensation and beam energy may occur, the results did not differ significantly from the described $Ar^+$ measurements.

## 4 CONCLUSION

The new high current LEBT satisfies the design expectations concerning full transmission, perfect beam matching to the RFQ phase space acceptance, and sharp-edged macropulses, even for an $Ar^+$ beam of 17 emA, which is 6 emA above the RFQ design intensity. In particular, the high degree of space charge compensation was confirmed. Since November 1999, as scheduled, the complete high current injector works reliable in routine operation [7].